# FIELD GALAXY EVOLUTION SINCE $z \sim 1$ FROM A SAMPLE OF QSO ABSORPTION–SELECTED GALAXIES

Charles C. Steidel[1,2,3,4]

MIT, Physics Department, Room 6-201, Cambridge, MA 02139

Mark Dickinson[1,2]

Space Telescope Science Institute, 3700 San Martin Drive, Baltimore, MD 21218

S. E. Persson

Observatories of the Carnegie Institute of Washington, 813 Santa Barbara Street, Pasadena, CA 91101

## ABSTRACT

We present some of the results of a large survey aimed at establishing the properties of galaxies selected by their having produced detectable Mg II $\lambda\lambda2796, 2803$ absorption in the spectra of background QSOs. The present sample covers the redshift range $0.2 \leq z \leq 1.0$, with $\langle z \rangle = 0.65$. From an extensive program of optical and near-IR imaging and optical spectroscopy, we find that the galaxies appear to be similar to normal galaxies at the present epoch, ranging from late–type spiral galaxies to those whose spectra and colors resemble present–day ellipticals. Contrary to some faint field galaxy samples selected using different criteria, over the redshift range observed we find no evidence for significant evolution in rest–frame $B - K$ color, space density, or (rest–frame $B$ or $K$) luminosity. The "average" Mg II absorbing galaxy appears to be consistent with a normal $0.7L_B^*$ Sb galaxy having a roughly constant star formation rate since $z \sim 1$, although galaxies spanning a range of a factor of $\sim 70$ in luminosity are found in the absorber sample. The diffuse gas cross-section selection imposed by studies of this kind appears to be biased *against* the relatively underluminous, blue galaxies which apparently dominate the number counts at faint magnitudes. However, essentially all "normal" field galaxies, independent of spectroscopic type, appear to be potential QSO absorbers.

*Subject headings:* galaxies: evolution–galaxies: distances and redshifts–quasars: absorption lines

[1] Visiting Astronomer, Kitt Peak National Observatory

[2] also Department of Astronomy, University of California, Berkeley, CA 94720

[3] Alfred P. Sloan Foundation Fellow

[4] NSF Young Investigator

## 1. Introduction

The study of the evolution of normal field galaxies at relatively high redshift has been revolutionized by the new generation of very efficient multiplexing faint object spectrographs. The samples chosen for study are selected on the basis of apparent magnitude in some particular observed pass-band (e.g., $B$, $I$, $K$), and are observed in one or a few small areas of sky in order to take advantage of the multi-object capabilities of the instruments (e.g., Broadhurst $et$ $al.$ 1988, Colless $et$ $al.$ 1989, Lilly 1993, Songaila $et$ $al.$ 1994.) Inherent limitations in this type of approach are that the identification of galaxies beyond a redshift of $\sim 1$ becomes increasingly difficult because the spectral features that are depended upon as redshift indicators (i.e., [OII] $\lambda3727$, Ca II K, H) begin to be redshifted beyond optical wavelengths, and one enters a regime in which there are essentially no conspicuous spectral features within the optical window for a large range of redshift. In addition, objects with "normal" luminosity become so faint that even if one can identify a redshift, one can hope for little else in the way of detailed information on the nature of the objects in question. Our alternative approach of using galaxies selected by their having produced detectable absorption lines in the spectra of background QSOs is motivated by the following (see also Steidel 1993b): 1) Normal objects can be followed to almost arbitrarily high redshift using identical and well-defined (rest–frame) selection criteria with well established statistics as a function of redshift 2) The selection criterion (essentially, gas cross-section) is not biased in principle by galaxy surface brightness, color, or luminosity– driven selection effects. The "completeness" of the sample is very well-defined–the existence of the object at the specific redshift is known $a$ $priori$, and so one can essentially search until the galaxy is found. 3) The objects are selected from a large number of (presumably) random lines of sight in all directions, hence biases introduced by large scale structures seen to be present in the pencil–beam redshift surveys are effectively "smoothed out". 4) One has $a$ $priori$ knowledge of the galaxy redshifts, so that if trends can be well-established at moderate redshift, the resulting predictions can be applied at much higher redshift so that direct spectroscopy (by far the most time–consuming aspect of any study of distant galaxies) becomes unnecessary for obtaining significant results on the evolution of the galaxy population. 5) The availability of a relatively bright QSO behind every galaxy in the sample allows, in principle, access to very detailed information such as chemical abundances, kinematics on scales of km s$^{-1}$, and gas–phase physical conditions through high resolution spectroscopy of the QSOs. Again, the detail to which one has access is largely independent of redshift and can be extended to the redshifts of the most distant QSOs.

In this letter, we present some of the results of our survey of absorption–selected galaxies that are particularly relevant to our understanding of the nature and evolution of normal galaxies since $z \sim 1$. Some of the preliminary results of this survey were given in Steidel and Dickinson (1992), and Steidel (1993a,b). The complete results and data will be published elsewhere (Steidel, Dickinson, & Persson 1995).

## 2. Sample Selection and Observations

We have selected our galaxy sample principally from two large, homogeneous absorption line surveys for Mg II $\lambda\lambda 2796, 2803$ doublets (Steidel & Sargent 1992; Sargent, Steidel, & Boksenberg 1988). This particular transition is chosen because it is observable from the ground over a very large range of redshifts ($0.2 \leq z \leq 2.2$), the ion is astrophysically abundant, and because it has been seen to be a very close tracer of optically thick H I gas (cf. Bergeron and Stasinska 1986). Selecting Mg II doublets with rest–frame equivalent widths $W_0 > 0.3$ Å provides a very nearly complete sample with a limiting H I column density of $\sim 10^{17}$ cm$^{-2}$ (see, e.g., Steidel 1992). For this reason, we often refer to our sample as being "gas cross-section selected"; it is not biased by the overall metallicity in the gas, which can range anywhere from solar to a few thousandths of solar and still remain consistent with the observed Mg II line strengths. The present sample consists of 58 galaxies along 48 different lines of sight, with a redshift range $0.2 \leq z \leq 1.0$ and $\langle z \rangle = 0.65$.

The observations for the survey of absorption–selected galaxies were made over the course of the last 4 years at a number of observatories. The optical imaging was carried out at the Kitt Peak National Observatory's 2.1m and 4m telescopes, and the 2.4m Hiltner telescope of the Michigan-Dartmouth-MIT Observatory. Most of the faint galaxy spectroscopy was completed using the Lick 3.0m Shane telescope. The near-IR imaging was done using NICMOS 3 HgCdTe arrays on the Kitt Peak 4m Mayall telescope and the 2.5m DuPont telescope at the Las Campanas Observatory. Details of the observations and data reduction procedures are presented elsewhere (Steidel *et al.* 1995).

The identification process is similar to that used by Bergeron and Boissé (1991), although we have placed more emphasis on obtaining high quality images across a wide range of wavelength. We first obtain deep optical images of the each field, and great care is taken in subtracting off the QSO light profile to reveal resolved objects which are very close to the QSO position. In practice, this is absolutely necessary for a complete census of galaxies near the line of sight even with our typical image quality of 0.9-1.0″ FWHM. The choice of optical filters to use for each field was motivated by the *a priori* knowledge of the galaxy redshift(s). The objective was to be able to obtain a measure of each galaxy's *rest–frame B* magnitude, so that the sample might be compared directly with "local" galaxy samples. Similarly, the $K$ band observations were designed to provide a measure which is much more directly relatable to a total stellar mass than the $U$ and $B$ (rest) magnitudes covered in the optical images. The k-corrections in this passband are quite small and relatively independent of the galaxy type; the $K$ band observations become increasingly important in extending this type of survey to higher redshift (cf. Aragon-Salamanca *et al.* 1994).

After careful PSF subtraction, we have been able to identify the absorbing galaxy in every line of sight in the sample; about 70% of the galaxies have been confirmed spectroscopically to have the same redshift as the absorption system, and in the remaining 30% a clear candidate, having intrinsic luminosity, color, and impact parameter fully consistent with the confirmed cases, has been identified. In addition to the absorber fields, we have also imaged ∼25 "control" fields: lines of sight taken from the same QSO absorption line surveys, but where *no* Mg II absorption is observed over the redshift range of interest. From both the absorber and the control fields, we know that incidences of "interloper" galaxies, i.e. galaxies at distances from the line of sight consistent with the absorbers but *not* producing detectable absorption, are very rare (see §3.2 below). There are clear correlations between the identified galaxies and the absorption line properties which suggest that the absorption is directly associated with the identified objects, rather than with an unseen companion (see Steidel 1993b, Steidel *et al.* 1995 for a complete discussion).

## 3.   PRINCIPAL RESULTS

### 3.1.   Galaxy Colors

One of the most surprising results of the survey has been the apparent lack of evolution in many of the properties for which one might *expect* evolution. As can be seen in Figure 1a, the *rest frame* $B - K$ color as a function of redshift for the galaxies basically encompasses the range expected for normal galaxy spectroscopic types with no evolution, from late type spirals to the red envelope of unevolved ellipticals. The mean color remains that of a *present day* mid–type spiral galaxy over the whole observed range. Clearly there has been no drastic change in the optical/IR color of the objects at least over the range encompassed by our observations, and the mean color corresponds closely to what might be expected on the basis of the local distribution of "normal" field galaxies. Minimal evolution in optical/IR color is consistent with a galaxy population having a roughly constant star formation rate (SFR) with time (cf. Bruzual and Charlot 1993).

### 3.2.   Luminosity Distribution

We find no evidence for significant evolution in the absolute $B$ or $K$ magnitudes as a function of redshift over the range spanned by our observations (see Fig. 1b). It is of course more difficult to compare the high redshift samples with the local samples from which luminosity functions have been constructed. However, the mean luminosities that we find for our absorption-selected objects are very much in line with what might be expected if the normal galaxy population of today were placed at $\langle z \rangle = 0.65$. This point is illustrated in Figure 3, where we show plots of conventional luminosity functions (with an arbitrary normalization) in rest–frame $B$ and $K$ as compared with some recent determinations from local samples of galaxies. The absorbing galaxy luminosity functions have been corrected for the observed dependence of gas cross-section on galaxy luminosity (and, hence, the relative volume probed by the absorption line survey at a given

galaxy luminosity:[5]

$$\sigma(L) \propto L^{0.4}$$

(see Steidel 1993, Steidel *et al.* 1995). Perhaps surprisingly, the gaseous size of the galaxies shows a much tighter relationship with $K$ luminosity than with $B$ luminosity, although the maximum likelihood dependence on luminosity has the same form.

The shape of the $B$-band luminosity function for the absorbing galaxies is striking–it looks much more like a Gaussian (parabola with the plotted axes) function than a Schechter (1976) function; on the other hand, the shape of the $K$-band luminosity function is in very good agreement with that of Mobasher *et al.* 1993, all the way down to galaxies that are only $\sim 0.05 L_K^*$ in luminosity. How can the two very different–looking luminosity functions be reconciled? An important clue comes from considering the local field galaxy luminosity function subdivided according to morphological type (Binggeli, Sandage, & Tammann 1988). Selecting only morphological types earlier than Sd (and excluding the dwarf ellipticals) results in a distribution very similar to that seen for the absorbers at $B$ in Figure 3, with the same "peak" in the luminosity function near $M_B = -20$ and the same "roll–off" toward fainter magnitudes. It is important to note that the faint end decline in the absorber $B$ luminosity function cannot be ascribed to incompleteness in the usual sense, as we have found the absorber in every case where we have looked.

There is a strong correlation between the rest–frame $B - K$ color and $M_K$ for the absorbing galaxies, in the sense that fainter galaxies have substantially bluer colors on average. No such correlation is apparent for $B - K$ versus $M_B$. The difference between the shapes of the $B$ and $K$ luminosity functions is a consequence of this correlation: the galaxies "missing" from the faint end of the $B$ luminosity function are intrinsically faint in the infrared but are very blue and thus apparently bright optically. They are essentially members of the "excess" population of so–called "faint blue galaxies" found in large numbers in the deep redshift surveys. Their nearest equivalent in the Binggeli *et al.* luminosity function (in luminosity and color, at least, if not in number density) would most likely be objects of very late morphological type.

In fact, the *only* galaxies which have been identified as true "interlopers" (i.e., those galaxies that are within the characteristic distance $R(L_B)$ of the absorbing galaxies but which do not produce detectable absorption) are both intrinsically faint at $K$ ($M_K \leq -22$) *and* very blue in their optical and optical/IR colors. In general, we find that if a galaxy has $M_K < -22$, it will be an absorber independent of its optical/IR color, whereas galaxies fainter than $M_K \approx -22$ apparently do not contribute substantially to the gas cross–section of the universe, and thus they are largely absent from our sample. At large redshifts, field galaxies that are this faint in absolute $K$ luminosity are also very blue (cf. Songaila $et\ al.$ 1994), so that they may contribute signficantly to the field galaxy $B$ luminosity function only 1.5 magnitudes fainter than $M_B^*$, yet be absent from our sample. In view of these arguments, we suggest that there may be a $morphological$ selection criterion in effect, and that gas cross-section appears to be driven more by galaxy mass (i.e., $L_K$) than by specific star formation rate.

---

[5]This correction is analogous to accounting for the differing volumes over which galaxies of a given absolute luminosity can be detected in a magnitude–limited survey, although it is a much shallower function of luminosity (i.e., $\propto L^{0.4}$ instead of $\propto L^{1.5}$). Thus, only small corrections are needed to go from the observed luminosity distribution to the true distribution, and one effectively samples a wide range of luminosities (and colors) independent of redshift, unlike surveys limited by apparent magnitude.

We have a direct measurement of the space density of the absorbers from a combination of the $dN/dz$ curve from the absorption line surveys (Steidel & Sargent 1992) and the observed impact parameters (allowing one to infer the total cross-section per galaxy of a given luminosity) of the galaxies from the QSO sightlines. The absorption line surveys have already shown that the Mg II systems at these equivalent width thresholds are consistent with no evolution in co-moving total gas cross-section over the whole range $0.2 \leq z \leq 2.2$ (Steidel & Sargent 1992); from our galaxy identifications, we find no change in the distribution of observed impact parameters as a function of redshift over the range of our survey. These two facts imply that the space density of the galaxies which produce absorption lines has not changed significantly (i.e., by no more than $\sim 30\%$) over the range of our survey. If we use the $K-$band luminosity distribution and our empirical luminosity–cross-section scaling relation to obtain a normalization for a conventional Schechter (1976) luminosity function, we find

$$\Phi^*(K) = 3.0 \pm 0.7 \times 10^{-2}\ \ \mathrm{Mpc}^{-3}$$

(for $H_0 = 100$ km s$^{-1}$ Mpc$^{-1}$). This number may then be compared directly with the local field samples– it is about a factor of 2 larger than the $B-$band normalization of Loveday $et\ al.$ (1992), and a factor of $\sim 4$ larger than the very low normalization found by Mobasher $et\ al.$ for the local $K-$band sample.

## 4.  DISCUSSION

While a complete discussion of the implications of our survey will be presented elsewhere (Steidel $et\ al.$ 1995), we summarize a few of the key points raised above. First, gas cross-section selection appears to pick out galaxies which on average have the spectroscopic characteristics of mid–type spiral galaxies, but span the range of colors expected for all (unevolved) normal galaxy spectroscopic types. It may come as some surprise that even apparently quiescent galaxies with little or no current star formation (based on their spectroscopic properties) appear to have substantial envelopes of diffuse gas capable of producing detectable absorption lines, and that the nature of these envelopes does not appear to correlate with the current star formation rate, but is more a function of overall galaxy mass. We find no evidence for any significant evolution in optical/IR color, $B$ or $K$ luminosity, or space density of absorbing galaxies over the range of redshift spanned by the survey. The normalization required for the population of absorbing galaxies is $\sim 2$ times higher than the most recent local field galaxy luminosity functions. Our results suggest that if this difference in normalization (relative to zero redshift) is taken at face value, then all of the changes in the galaxy population must have occurred at redshifts smaller than $\sim 0.3$. Specifically, merging and/or luminosity evolution of the normal, relatively luminous galaxies selected by gas cross-section cannot be important in the redshift range $0.3 \leq z \leq 1.0$. The relatively faint, very blue galaxies which make significant contributions to the faint galaxy counts and which are present in large numbers in the apparent magnitude selected redshift surveys (e.g. Broadhurst $et$ $al$ 1988, Colless $et$ $al.$ 1989, Songaila $et$ $al$ 1994, Cowie $et$ $al$ 1992) apparently do not possess substantial gas cross-section, as they are conspicuously absent from our survey. It is this population of objects which poses problems to models of galaxy evolution; when the population of "small" blue galaxies is excluded (as they are using our selection criterion), there is apparently nothing spectacular happening at all for the remaining "normal" galaxy population.

Because of the very specific predictions that can be made on the basis of our $z \leq 1$ sample, higher redshift surveys can be feasibly pursued without the benefit of follow-up galaxy spectroscopy, i.e. using deep optical and IR imaging alone. The predicted colors and magnitudes of absorbers at $z \sim 1.5$ are such that complete spectroscopy in the optical would prove almost impossible using current technology (cf. Songaila $et$ $al.$ 1994). We are in the process of completing a new sample of absorbers in the redshift range $1.0 \leq z \leq 1.6$ (Steidel & Dickinson 1995), a regime over which the properties of normal galaxies are essentially unexplored. In addition, the advent of 8–10m telescopes will allow wholesale high resolution spectroscopy along the same lines of sight in order to explore the detailed physics of the same galaxies, all as a function of redshift.

We would like to thank Max Pettini for very constructive comments on an earlier draft of this paper.

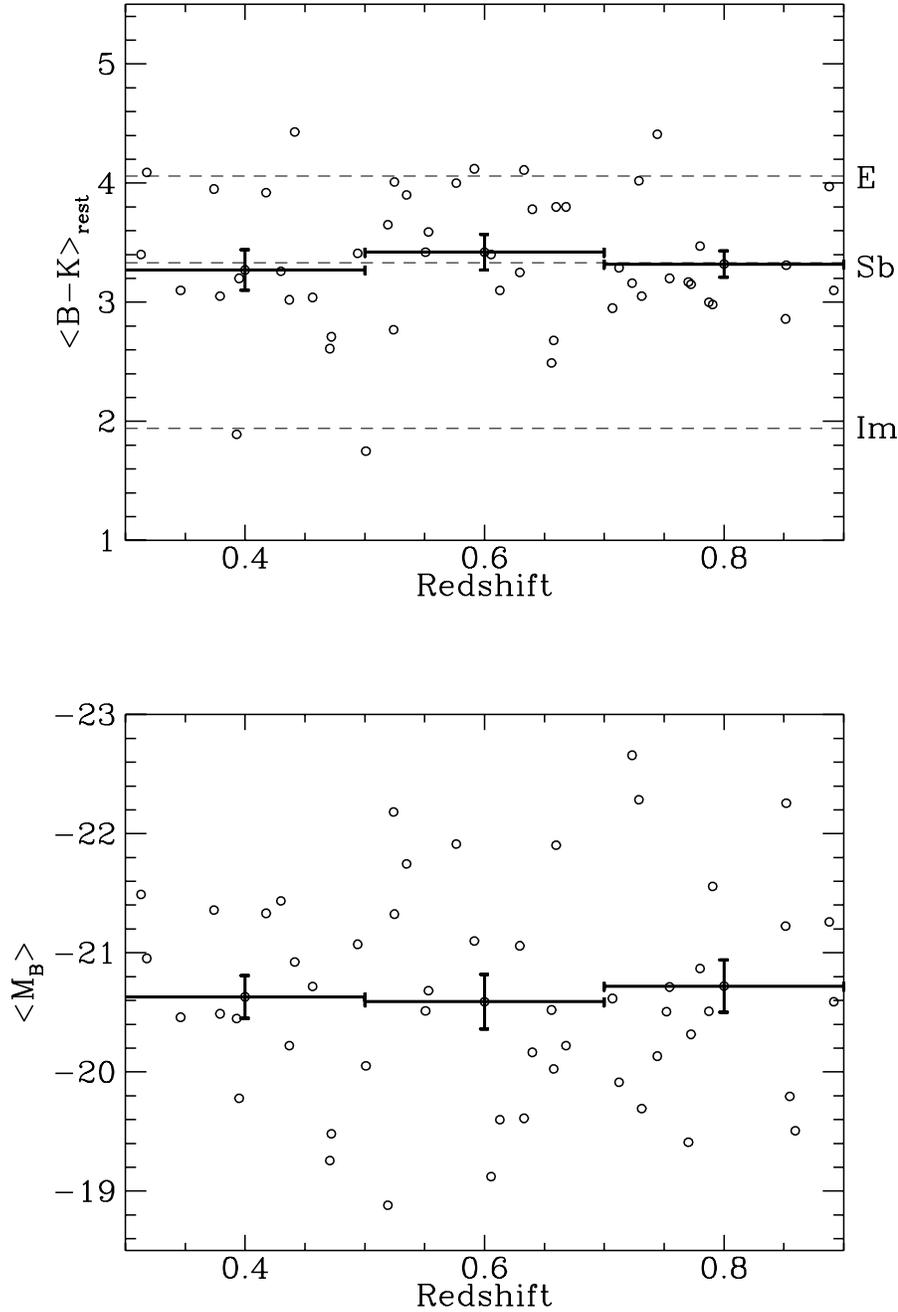

Fig. 1.— a) The rest–frame $B - K$ color of the absorbing galaxies versus redshift (open symbols), with the mean values indicated using error bars in 3 redshift bins. The dashed horizontal lines correspond to the rest–frame colors of Im, Sb, and E (in order of increasing $B - K$) galaxy models (Bruzual and Charlot 1993). Note that the mean color is almost exactly that of the Sb model SED. b) The rest–frame $B$ luminosity of the galaxies as a function of redshift, for $H_0 = 50$ km s$^{-1}$ Mpc$^{-1}$ and $q_0 = 0.05$. There is no significant evolution in either the color or luminosity of the absorbing galaxies over the range observed.

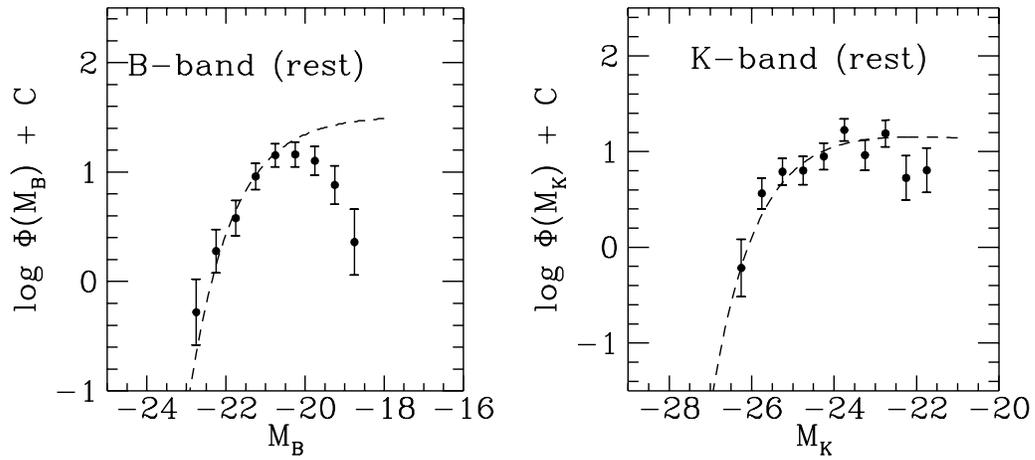

Fig. 2.— Plots of the rest–frame *B* and *K* luminosity functions (with arbitrary normalization) for the absorbing galaxies. The local luminosity functions (dashed curves) are taken from Loveday *et al.* 1992 (*B*) and Mobasher *et al.* 1993 (*K*).